\documentclass[twocolumn,showpacs,preprintnumbers,amsmath,amssymb,aps]{revtex4}
\usepackage[english]{babel}
\usepackage[cp1250]{inputenc}
\usepackage{fancyhdr}
\usepackage{graphicx}
\usepackage{color}
\usepackage{amsmath}
\usepackage{amssymb}
\usepackage{hyperref}
\usepackage{epsfig}
\usepackage{amsthm}
\usepackage{graphicx}
\usepackage{dcolumn}
\usepackage{amsmath,amsthm,latexsym,amssymb,amsfonts,epsfig}
\usepackage{fontenc,dsfont,layout}
\newcommand{\be}{\begin{equation}}
\newcommand{\ee}{\end{equation}}
\newcommand{\ba}{\begin{eqnarray}}
\newcommand{\ea}{\end{eqnarray}}

\begin{document}
\preprint{preprint}

\title{Dynamic transition between Fresnel and Fraunhofer diffraction patterns - a lecture experiment}

\author{Maciej Lisicki}
\email{mklis@fuw.edu.pl}

\author{Ludmiła Buller} 
\author{Michał Oszmaniec}
\email{moszm@okwf.fuw.edu.pl}
\author{Krzysztof Wójtowicz}

\affiliation{
Faculty of Physics, Warsaw University, Hoża 69, 00-681 Warsaw, Poland
}

\date{\today}

 \begin{abstract} 
A simple method for presenting a dynamic transition between Fresnel and Fraunhofer diffraction zones is considered. Experiments are conducted on different apertures and diffraction patterns are photographed at various distances between the screen and the aperture. A diverging lens is introduced into the experimental setup to provide enlarged Fresnel diffraction patterns. Fresnel and Fraunhofer diffraction patterns and dynamic transition between them can be easily obtained on a distance of few meters, what gives an opportunity to use our setup as a lecture experiment. Photographs of transition for square aperture are shown and discussed. 
\end{abstract} 	
 \maketitle
%
%
%
%
\section{Introduction}

Fresnel and Fraunhofer diffraction and shapes of diffraction patterns appear as an important part of any optics lecture, often accompanied by a presentation of the phenomenon, but this includes only Fresnel or Fraunhofer diffraction on common and simple apertures. It is mainly because these patterns can be easily calculated, so that one can compare theoretical and experimental results\textsuperscript{\cite{Hecht},\cite{Pain},\cite{Crawford}}. 
Plenty of articles have been devoted to lecture demonstrations\textsuperscript{\cite{laserAJP}} or inexpensive student experiments on Fresnel\textsuperscript{\cite{studentAJP1},\cite{studentAJP2},\cite{studentAJP3}} or Fraunhofer\textsuperscript{\cite{triangleAJP},\cite{platesAJP}} diffraction comapring theoretical results with the obtained pattern.

Our aim is to show the rarely presented but very interesting effect: the dynamic transition between Fresnel and Fraunhofer diffraction zones. It can provide an insight in the changing pattern structure and breaking of the plane wave approximation when approaching to the aperture.

We also present a method for preparing high--quality inexpensive apertures, more precise but easily accessible version of method involving a laser printer\textsuperscript{\cite{laserAJP},\cite{laser}}. 
\section{Fresnel and Fraunhofer diffraction zones}
 Diffraction can be classified as Fresnel or Fraunhofer (also called far field diffraction). In Fraunhofer diffraction the distance between the aperture and the screen is large enough to treat the wavefront as planar. In Fresnel diffraction the wavefront has to be treated as a curved surface.
 
 A simple criterion\textsuperscript{\cite{Crawford}} may be formulated to assess the these diffraction approximations. We consider a monochromatic light source and  assume that the phase of the incident wave is the same at each point on the aperture. Denoting the characteristic size of the aperture by $d$ (i.e. its diameter), considering a point at a distance $L$ from the aperture and using light of wavelength $\lambda$, the diffraction type can be described by a dimensionless parameter $\alpha$:
\be
\alpha=\frac{\lambda L}{d^2}
\ee
Generally, when $\alpha\gg 1$, we can use the Fraunhofer approximation. When $\alpha\leq 1$, we have to use Fresnel approximation. For example, with $\alpha=1$, $\lambda=650$ nm and an aperture with diameter of $d=1$ mm, we can estimate the distance as $L\approx1.5$ m. Thus the screen should be perhaps $5L\approx 7.5$ m from the aperture to get a proper Fraunhofer diffraction pattern. Obviously such a large distance is difficult to achieve in a lecture hall or classroom. On the other hand, Fresnel diffraction can be observed for $\alpha\leq 1$, so observing both types of diffraction on the same aperture may be complicated - the Fresnel image is very small and therefore it cannot be properly demonstrated. It is even more difficult to show the transition between them over such a big distance - one would have to move the aperture or the screen, what may cause a considerable loss in picture quality.

\begin{figure*}
\includegraphics[width=16cm]{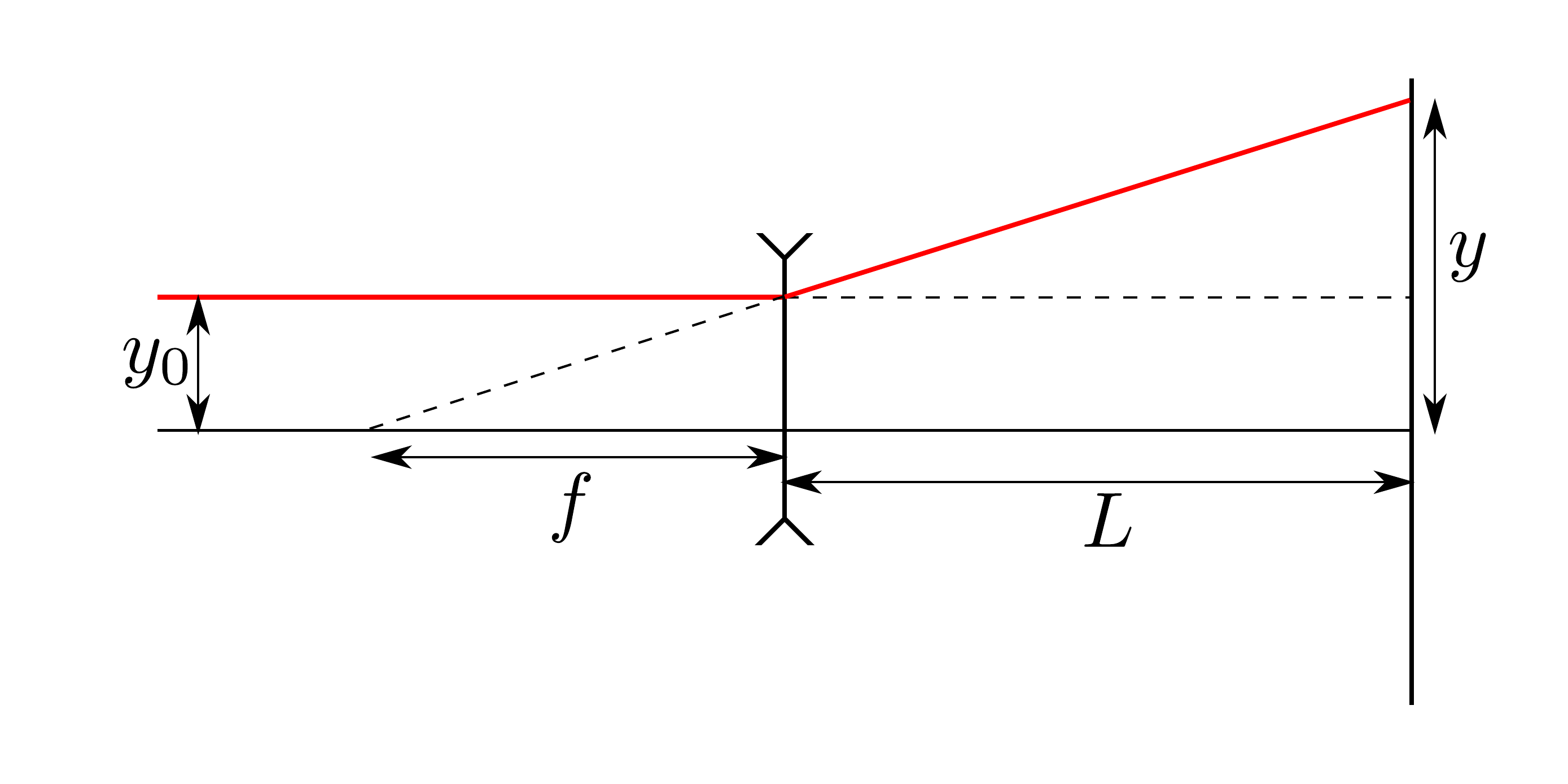}
\caption{\label{geo} Geometry of the setup.}
\end{figure*}

Let us now consider what will be the qualitative influence of placing a diverging lens between the aperture and the screen (see fig. \ref{geo}). In order to do so we trace a paraxial ray parallel to the optical axis. Note that after passing through the lens the ray will follow the path shown in the figure. If we assume that the lens is thin and that it is justifiable to use approximations of geometric optics, it is possible to derive an expression for magnification rate $M$ as 
\be
M=\frac{y}{y_0}=\Bigl(1+\frac{L}{|f|}\Bigr).
\ee
Where: $f$ - focal length of the lens, $L$ - distance between the lens and the screen, $y$, $y_0$ - distances from the ray to the optical axis on the screen and before coming through the lens, respectively. $M$  obviously exceeds 1. Although the same reasoning cannot be repeated for any arbitrary rays coming form the aperture, it is clear that the introduced diverging lens acts as a projector, effectively magnifying the pattern created on its surface. Rays that initially were close to the optical axis will be further from it when they reach the screen.

Hence, by introducing a diverging lens into the experimental setup, we can get larger images from a small aperture in Fresnel regime. This eliminates the problem of long distances and makes it possible to present dynamic change of the pattern while moving the lens between the aperture and the screen.

%
%
%
%
\section{Experimental setup}
\begin{figure*}
\includegraphics[width=16cm]{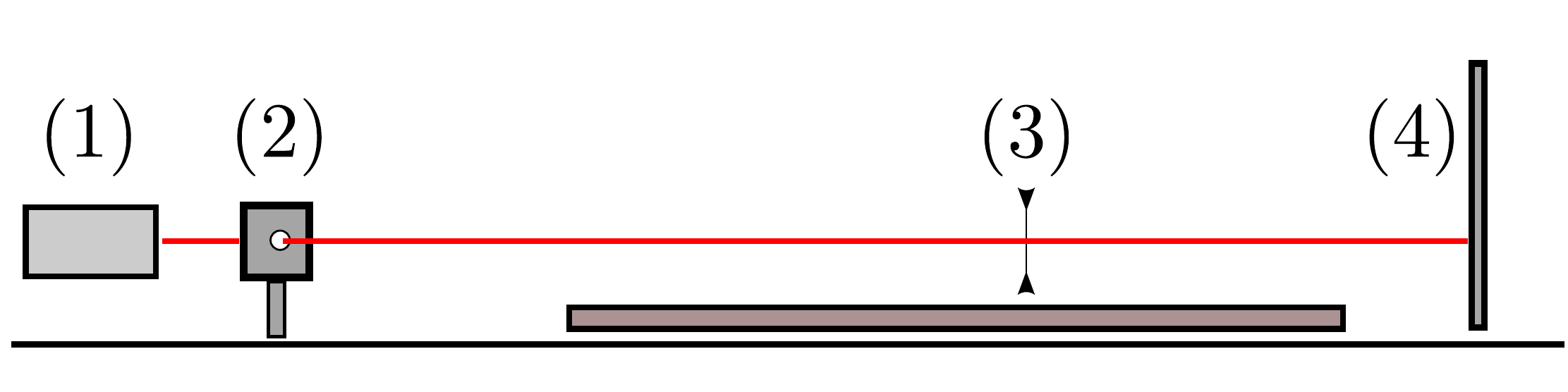}
\caption{\label{ukl} Scheme of the experimental setup.}
\end{figure*}

The experimental setup (fig. \ref{ukl}) consisted of He-Ne laser (1) which was used as a monochromatic light source (with wavelength $\lambda=650$ nm), an aperture (2), the screen (4), and a movable diverging lens (3) placed between the aperture and the screen. The screen was made of semi--transparent parchment, which allowed to observe the back of the screen. It appears to be the best way to observe the pattern and take photographs from behind because then light goes straight into the camera (or human eyes) and the image is symmetrical.  Any error in the screen position or aperture adjustment would result in disruption of the image. The main difference between our setup and the standard one is the use of a movable diverging lens. That allows to observe a transition between the two types of diffraction without moving the screen or the aperture and ensures good quality images. Thus, relatively small apertures ($d\approx1$ mm) could be used in the experiment, that shifts the Fraunhofer zone to closer distances. On the other hand, Fresnel pattern was small, but this problem has been solved by using the diverging lens to show magnified image on the screen. The size was larger, although the brightness was lower. We have used distances of about 3-5 meters. The lens had to be relatively large. We have used one of focal length $f=-30$ cm and diameter of about 10 cm.

One has to pay special attention to the preparation of apertures. For homemade good-quality apertures one can use a laser printer\textsuperscript{\cite{laserAJP},\cite{laser}} but for a more complicated shapes the printer resolution is insufficient. Our apertures were obtained by exposing the aperture image (prepared in a vector graphics programme) on a transparent foil with high resolution (in our case it was 4000 dpi). The foil can be exposed in a photo laboratory, so the aperture production is easily accessible and inexpensive. The aperutre quality is considerably higer and computers provide us with a tool to construct new apertures by varying their shapes and sizes.

In the context of showing the experiment to a wider audience the best way to demonstrate the transition is to put the video camera behind the screen and connect it to a multimedia beamer. It provides big and sharp images, what makes them accurate for demonstrating purpose.

%
%
%
%
\section{Diffraction images}

\begin{figure*}
\includegraphics[width=16cm]{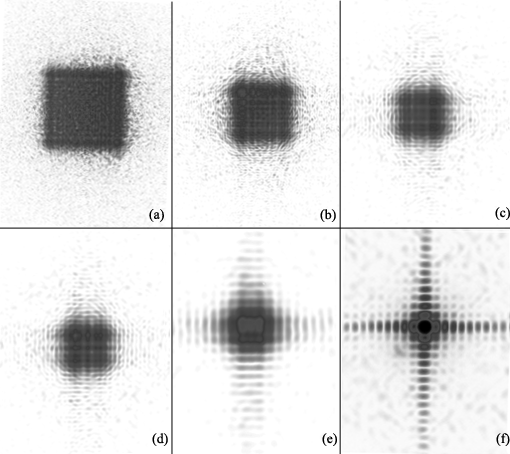}
\caption{\label{diff} Successive photographs of transition between Fresnel and Fraunhofer diffraction patterns.}
\end{figure*}

The successive photographs \ref{diff}(a)-\ref{diff}(f) present images of transition between Fresnel zone and quasi-Fraunhofer diffraction pattern for a square aperture. It is not an exact Fraunhofer image (for this, we would need bigger distances) but the image obtained at maximal possible distance does not differ significantly from the far-field image. That has been checked by placing the screen even further. The images were obtained by moving the lens from the aperture towards the screen. 

The photographs have inverted colours to make the structures more visible, so the brighter areas in the photographs correspond to the darker zones in the real image.
The photograph \ref{diff}(a) presents a typical Fresnel diffraction image\textsuperscript{\cite{Hecht}}. The grid-like structure inside a square is due to the bending of light rays on the borders of the aperture. There are lots of local minimums and maximums of light intensity due to constructive and destructive interference inside the square shape. Light intensity rapidly decreases outside the square. 

In the photograph \ref{diff}(b) the number of gratings is smaller. They are wider and we can also observe that the structure of the image changes - it still shows the symmetries, but the light intensity outside the square does not decrease rapidly. 

In the image \ref{diff}(c) we can observe the formation of a cross. Image \ref{diff}(d) shows clearer cross-like structure, both outside and inside. We can clearly see that two bright stripes get closer to merge completely in image \ref{diff}(e). 

The blurred structure in the middle of the picture is due to the fact that the centre of the image is much brighter than the other regions and during a long exposure time it receives much more light. We are now two-thirds of a way from the aperture to the screen. The limbs are clearly visible and we can also see their subtle structure - fringes. Placing the lens very close to the screen or completely removing it, we get an image similar to \ref{diff}(f). 

\begin{figure*}
\includegraphics[width=16cm]{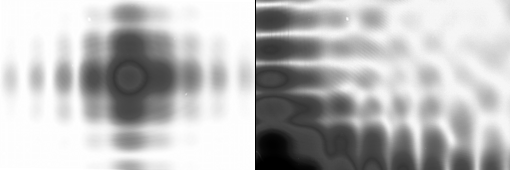}
\caption{\label{diff2} Subtle structure of the inter-limbs regions - beautiful example of Fraunhofer diffraction image.}
\end{figure*}

This is a classical Fraunhofer diffraction image. The limbs consist of clear fringes, and their intensity does not decrease rapidly as we recede from the central point. Also the regions between the limbs have an interesting structure, shown in figure \ref{diff2}. The whole image conserves the symmetries of the aperture. The structure of these regions is predicted by the theoretical description of Fraunhofer diffraction. We can look at this image as a product of two single slit diffraction patterns, as described in various textbooks\textsuperscript{\cite{Hecht},\cite{Pain}}.

All the photographs were taken with the use of a digital reflex camera. As the light intensity on the screen rapidly decreases with the distance between the aperture and the lens, photographs have been taken at various exposure times, up to 20 seconds. The images can be clearly seen by a naked eye but applying several exposure times allowed to investigate different parts of the image (the pattern is very bright in the middle but dark at the borders, so to see the darker regions we use the SLR camera).

\section{Conclusions}
We have presented an experimental setup that allows to show a dynamic transition between Fresnel and Fraunhofer diffraction. Thanks to the use of a diverging lens we were able to observe both types of diffraction on the same aperture in a relatively small setup, which makes it particularly useful for lecture demonstrations. 

Moreover, making new apertures is easy and available to almost everyone. It can be an interesting experiment to investigate transition patterns for common shape apertures, like slits or circular holes, as well as for more complicated ones. One can also consider symmetries of diffraction images and their connection with the geometry of apertures.

The presented images are only to show that it is possible to obtain high-quality diffraction images in school conditions, but the real transition is much more attractive to see. One can observe the dynamic changes of structure, the exact process of image formation.

More of our diffraction images obtained using the described method, including diffraction on simple shapes (ie. multiple circular dots in various configurations) and also fractal apertures like Sierpiński gasket or Koch curve\textsuperscript{\cite{fraktal}}, are available on the Internet\textsuperscript{\cite{WWW}}.

\begin{acknowledgments}
We would like to acknowledge M.Sc. Stanisław Lipiński from XIV Stanisław Staszic High School in Warsaw for fruitful discussions and advice in preparing the experiments, Ph.D. Piotr Kossacki, Ph.D. Piotr Szymczak and Ph.D. Przemek Olbratowski from Warsaw University for consultations and Maciej Zielenkiewicz for help in making the apertures.
\end{acknowledgments}
%
%
%
%

\begin{thebibliography}{9}

\bibitem{Hecht} {E. Hecht, \textit{Optics}, 4th ed. (Addison Wesley, 2002), Fraunhofer diffraction on a square aperture pp. 464--467, Fresnel diffraction on a square p. 499.}

\bibitem{Pain} {H. J. Pain, \textit{Physics of Vibrations and Waves}, 6th ed. (Wiley, 2005), pp. 377--386.}

\bibitem{Crawford} {F. S. Crawford, \textit{Waves}, Berkeley Physics Course, (McGraw - Hill, 1968), pp. 457--488.}

\bibitem{laserAJP} {J. van der Gacht, \textit{Simple method for demonstrating Fraunhofer Diffraction}, Am. J. Phys \textbf{62}, pp. 934--937, (1994).}

\bibitem{studentAJP1} {P. A. Young, \textit{Student experiment in Fresnel Diffraction}, Am. J. Phys \textbf{32}, 367--369 (1964).}

\bibitem{studentAJP2} {L. A. Sanderman, R. S. Bradford, \textit{A Simple Fresnel Diffraction Experiment}, Am. J. Phys \textbf{17}, 514 (1949).}

\bibitem{studentAJP3} {A. L. Moen, D. L. Vander Meulen, \textit{Fresnel Diffraction using a He--Ne Gas Laser}, Am. J. Phys \textbf{38}, 1095--1097 (1970).}

\bibitem{triangleAJP} {M. J. Moloney, W. Meeks, \textit{Experiment in Fraunhofer Diffraction Using a Triangular Aperture}, Am. J. Phys. \textbf{42}, 696--698 (1974).}

\bibitem{platesAJP} {R. B. Hoover, \textit{Diffraction Plates for Classroom Demonstrations}, Am. J. Phys. \textbf{37}, 871--876 (1969).}

\bibitem{laser} {S. J. Van Hook, \textit{Inquiry with Laser Printer Diffraction Gratings}, Phys. Teach. \textbf{45}, pp. 340--343 (2007).}

\bibitem{fraktal} {J. Uozumi, K.-E. Peiponen, M. Savolainen, R. Silvennoinen and T. Asakura, \textit{Demonstration of diffraction by fractals}, Am. J. Phys \textbf{62}, 283--285 (1994).}

\bibitem{WWW} {A collection of Fraunhofer and Fresnel diffraction images, obtained at varoius distances from the aperture using the method described in the article: \texttt{www.fuw.edu.pl/\textasciitilde mklis/diff.html}.}

\end{thebibliography}
\end{document}